\documentclass[preprint2]{aastex}

\slugcomment{Not to appear in Nonlearned J., 45.}

\shorttitle{Carbon-enriched metal poor stars}
\shortauthors{Djorgovski et al.}

\usepackage{natbib}
\begin{document}


\title{Formation of carbon-enhanced metal-poor stars in the presence of far ultraviolet radiation}


\author{S. Bovino\altaffilmark{1}, T. Grassi\altaffilmark{2,3}, D. R. G. Schleicher\altaffilmark{1}, and M. A. Latif\altaffilmark{1}}
\affil{$^1$Institut f\"ur Astrophysik Georg-August-Universit\"at, Friedrich-Hund Platz 1, 37077 G\"ottingen, Germany}
\affil{$^2$Centre for Star and Planet Formation, Natural History Museum of Denmark, \O ster Voldgade 5-7, 1350 Copenhagen, Denmark}
\affil{$^3$Niels Bohr Institute, University of Copenhagen, Juliane Maries Vej 30, 2100 Copenhagen, Denmark}
\email{sbovino@astro.physik.uni-goettingen.de}






\begin{abstract}
Recent discoveries of carbon-enhanced metal-poor stars like SMSS J031300.36-670839.3 provide increasing observational insights into the formation conditions of the first second-generation stars in the Universe, reflecting the chemical conditions after the first supernova explosion. Here, we present the first cosmological simulations with a detailed chemical network including primordial species as well as C, C$^+$, O, O$^+$, Si, Si$^+$, and Si$^{2+}$ following the formation of carbon-enhanced metal poor stars. The presence of background UV flux delays the collapse from $z=21$ to $z=15$ and cool the gas down to the CMB temperature for a metallicity of Z/Z$_\odot$=10$^{-3}$. This can potentially lead to the formation of lower mass stars. Overall, we find that the metals have a stronger effect on the collapse than the radiation, yielding a comparable thermal structure for large variations in the radiative background. We further find that radiative backgrounds are not able to delay the collapse for Z/Z$_\odot$~=~10$^{-2}$ or a carbon abundance as in SMSS J031300.36-670839.3.
\end{abstract}





\section{Introduction}
Recent discoveries of extremely metal-poor stars \citep{Christlieb2002,Frebel2005,Norris2007,Keller2014} raised the question from which mechanism and under which conditions they were formed.
These stars present very low iron abundances [Fe/H]~$<$~-4 but an anomalous richness in carbon and are considered crucial to provide a physical insight on the possible environments where the first stars were formed.
A distinct class of carbon-enhanced metal-poor (CEMP) stars exists and different mechanisms for their formation have been proposed \citep{Umeda2003}.

Stars like SMSS J031300.36-670839.3 are thought to be formed from a low-energy ($\sim 10^{51}$ erg) type II supernova of a primordial massive star ranging between 40-60 M$_\odot$ \citep{Keller2014}. Violent pair-instability supernovae (PISNe) are in fact likely to disrupt the hosting halo inhibiting the formation of a second generation of stars \citep{Greif2007,Cooke2014,Seifried2014}. Low-explosion energy black-hole forming events \citep{Umeda2003}  need to be invoked to explain the abundance pattern observed in CEMP stars \citep{Keller2014}. During such events metals heavier than iron are trapped inside the black hole because of the larger degrees of fallback, whilst lighter elements which reside in the outer region are dispersed during the explosion. This scenario has been recently investigated by \citet{Cooke2014} through detailed nucleosynthesis calculations. They also suggested that the seeds of CEMP stars should have been relatively low mass halos of a few 10$^6$ M$_\odot$. 

Simplified theoretical models \citep{Frebel2007,Safranek2010,Ji2014} suggested possible conditions to form low mass CEMP stars that can be observed today by evaluating the amount of metals needed to induce cooling and subsequent fragmentation.
\citet{Salvadori2007,Salvadori2010} explored the metallicity distribution function  to probe the stellar population history of the Milky Way providing a critical metallicity of Z$_\mathrm{crit}$/Z$_\odot$ = 10$^{-4}$.
 Nevertheless, while the Population III star formation mode was the main focus of many hydrodynamical simulations, indicating typical masses between 10-300 M$_\odot$ \citep[][and references therein]{Hirano2014}, the transition between the Pop III - PopII star formation mode is still an ongoing research topic which needs to be accurately explored\footnote{A window of possible low-mass primordial star formation via fragmentation or HD cooling has also been explored by several authors \citep{ClarkGlover2011,Greif2012,Stacy2012,Prieto2013,Bovino2014mergers}.}.

Results from smoothed particle hydrodynamics (SPH) calculations have been reported by \citet{Bromm2001} which propose a critical metallicity Z$_\mathrm{crit}$/Z$_\odot$ = 10$^{-3.5}$ which regulates the transition between the high and low-mass star formation mode. This value was suggested to be unlikely by \citet{Jappsen2009} indicating that fragmentation processes mostly depend on the initial conditions. 
They explored conditions under which cooling is mainly regulated by fine-structure transitions of oxygen and carbon by employing an idealized Navarro-Frenk-White (NWF) density profile and a low-metallicity network for a minihalo of 2$\times$10$^6$ M$_\odot$.

Recently, \citet{Safranek2014} carried out a series of calculations starting from realistic cosmological initial conditions for an atomic cooling halo ($\sim$10$^7$ M$_\odot$) irradiated by a fixed UV flux (J$_{21}$ = 100). They included a comprehensive model for non-equilibrium chemistry and metal line cooling and followed the evolution of the halo by introducing sink particles.   
The predicted metallicity threshold given by \citet{Bromm2001} has been confirmed and the influence of metal line cooling on the final masses has been discussed. In particular they found that a metallicity of Z/Z$_\odot$~=~10$^{-3}$ is needed to reach the cosmic background temperature and to induce fragmentation, forming a stellar cluster of $\sim$ 1000 M$_\odot$. 

In this letter we explore the chemo-dynamical conditions of a metal-poor minihalo required to cool and undergo gravitational collapse to provide a possible site of CEMP star formation. 
We start our calculations from cosmological initial conditions including a comprehensive model for the non-equilibrium metal cooling and photochemical processes and vary both the metallicity of the gas as well as the UV radiation strength. 

In the following sections, we provide the details of our simulations, present the main results, and summarize our conclusions.
\section{Simulations setup}
To follow the collapse of typical minihalo from cosmological initial conditions we employ the cosmological hydrodynamics code \verb|ENZO|, version 2.3 \citep{Enzo2014}. \verb|ENZO| is based on an adaptive mesh refinement (AMR) method. It includes the split 3rd-order piece-wise parabolic (PPM) method for solving the hydrodynamical equations, while the dark matter component is modeled using the particle-mesh technique. Self-gravity is calculated via a multi-grid Poisson solver.
We first run a dark-matter only low-resolution simulation to study the evolution of the halo for a box of 300 kpc/h with a top grid resolution of 128$^3$ cells similarly to \citet{Bovino2013MNRAS,Bovino2014} and \citet{Latif2013}.
 The parameters for creating the initial conditions and the distribution of baryonic and dark matter components are taken from the WMAP seven year data \citep{Jarosik2011}. We select the most massive mini-halo of $\sim$1.4$\times10^5 $ M$_\odot$ and re-run our simulation with the box centered onto the selected mini-halo adding two additional nested grids for an effective resolution of 512$^3$.
From $z=99$ to $z=22$ we evolve the halo to reach a temperature of $\sim$10$^3$~K. We then inject metals which are initialized based on the metallicity. We allow 20 levels of refinement and 32 cells per Jeans length. 
Our refinement strategy is based on over-density, Jeans length, and particle mass and is applied during the course of the simulations to ensure that all physical processes are well resolved and the Truelove criterion \citep{Truelove1997,Federrath11} is fulfilled. 

\begin{figure}[!h]
\includegraphics[scale=.40]{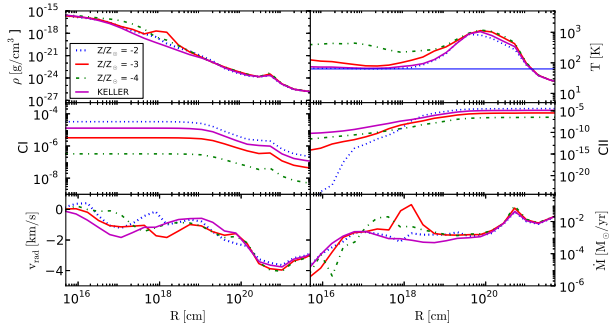}
\caption{Radial profiles of the averaged density (top left), temperature (top right), neutral and ionized carbon (middle panels), radial infall velocity (bottom left), and accretion rate (bottom right). 
Each curve represents a different metallicity run as sketched in the legend. The horizontal blue line represents the CMB temperature.}\label{figure1}
\end{figure}

\subsection{Chemistry}
The evolution of the gas enriched by metals and irradiated by UV background is regulated by the complex interplay between cooling, kinetics, and hydrodynamics. 
A standard approach to consider metal cooling in hydrodynamical simulations is to employ equilibrium \textsc{cloudy} tables making use of a general metallicity field \citep[e.g.][]{Smith2008,Smith2009}, or introducing equilibrium approximations for some of the species which evolve faster \citep{GloverJappsen2007,Jappsen2007ApJ}. Recently, \citet{Peters2014} explored metal cooling employing a parametrized equation
of state. 
To our knowledge a complete non-equilibrium approach including the coupling between kinetics and cooling for low-metallicity environments has never been implemented in \textsc{enzo} and only a few simulations have been performed with other codes \citep[see][]{GloverJappsen2007,Safranek2014}. In this study we employ the publicly available chemistry package \textsc{krome}\footnote{\url{www.kromepackage.org}} \citep{Grassi2014} to evolve 16 species: H, H$^+$, H$^-$, H$_2$, H$_2^+$, He, He$^+$, He$^{2+}$, C, C$^+$, Si, Si$^+$, Si$^{2+}$, O, O$^+$, and e$^-$, for a total of 44 reactions. We include photoionization and photoheating for C and Si, which are easily ionized for fluxes below the Lyman limit, with ionization thresholds of 8.15 eV and 11.26 eV, respectively. We integrate the cross sections provided by \citet{Verner1996} as described by \citet{Grassi2014} considering an optically thin gas. Photodissociation of H$_2$, H$_2^+$, H$^-$ photo-detachment, collisional dissociation by \citet{Martin1996}, and the self-shielding function from \citet{Wolcott2011sfh} are also employed \citep[see also][]{Latif2014}. We consider here a T$_*$ = 10$^4$ K soft spectrum. The following cooling/heating processes are included: H$_2$ cooling, atomic line cooling, chemical heating/cooling, and metal line cooling. The latter is evaluated on the fly solving the linear system for the individual metal excitation levels along with the rate equations integration, as discussed in \citet{GloverJappsen2007,Maio2007} and \citet{Grassi2014}. For additional details on the thermal processes implemented in \textsc{krome} we refer to \citet{Grassi2014}. 
To prevent the temperature from dropping below the CMB floor and since induced (de)-excitations are not included, we define the following effective cooling rate:
\begin{equation}
	\mathrm{\Lambda_{eff} = \Lambda_{metal} (T_{gas}) - \Lambda_{metal} (T_{CMB})}
\end{equation} 
as also reported by \citet{Jappsen2009a}.
We do not include deuterium chemistry as well as HD cooling which is easily photodissociated by a weak radiation background \citep{Wolcott2011MNRAS}. We note that all the chemical species are consistently advected to ensure conservation.
The network used here is publicly available with the package \textsc{krome} (react\_lowmetal). The reactions for metals are taken from \citet{GloverJappsen2007}, numbers 30-47, 56 and 58, while the primordial reactions are the standard reactions employed in \textsc{krome} \citep[see][table C1]{Grassi2014} with the only difference in the three-body formation rate that we take from the latest available data \citep{Forrey2013}.

The initialization of the individual metal species X is based on the following definition:
\begin{equation}
	\log_{10}(n_\mathrm{X}/n_\mathrm{H}) = \mathrm{Z/Z_\odot}+\log_{10}(n_\mathrm{X}/n_\mathrm{H})_\odot
\end{equation}
with Z/Z$_\odot$ being the logarithm of the metallicity expressed in terms of solar metallicity and $n_\mathrm{H}$ the total hydrogen nuclei. 
The solar abundances $(n_\mathrm{X}/n_\mathrm{H})_\odot$ are taken from \citet{Asplund2009}.
Only in one case we directly use the observed abundances for the CEMP star SMSS J031300.36-670839.3 \citep{Keller2014}.

\begin{figure*}
\includegraphics[scale=.55]{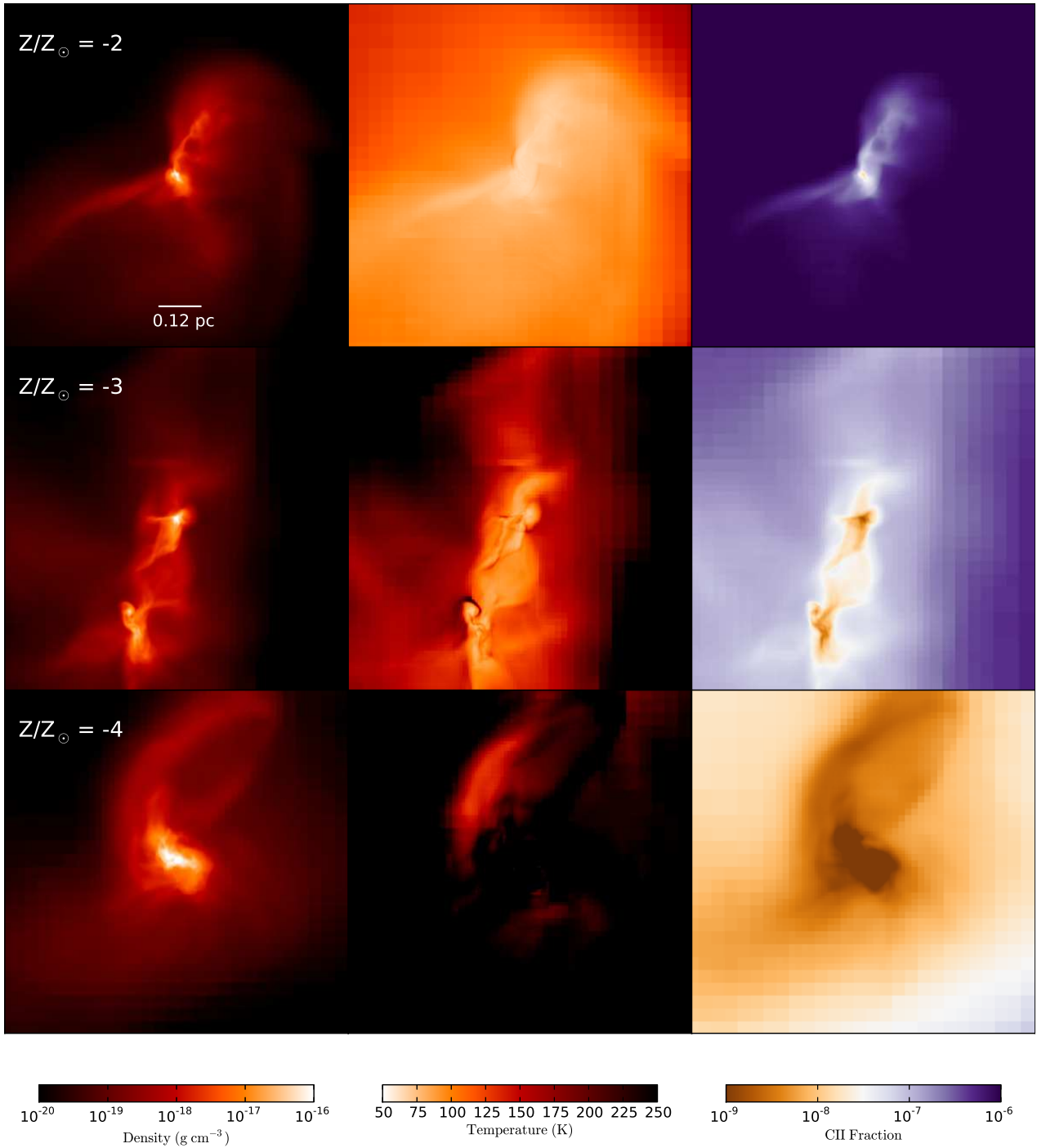}
\caption{Density, temperature, and CII projections along the y-axis at a scale of 1 pc, for three different metallicities as sketched in the plots.}\label{figure2}
\end{figure*}

\begin{figure*}
\includegraphics[scale=.8]{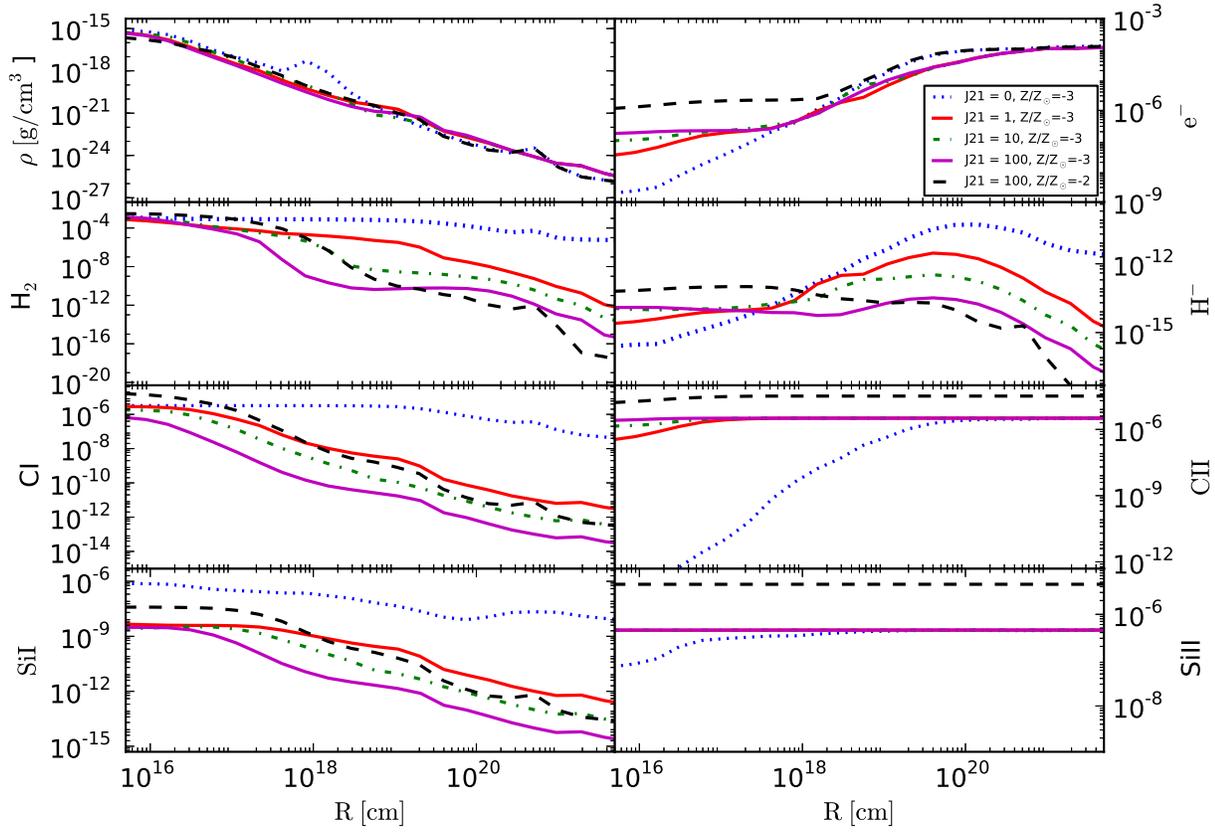}
\caption{Radial profiles of the density and the most important chemical species fractions: electrons (top right), H$_2$ and H$^-$ (middle top panels), CI and CII (middle bottom panels), and SiI and SiII (bottom panels). Different values of the flux strength in terms of J$_{21}$ and metallicities (Z/Z$_\odot$) are reported according to the legend.}\label{figure3}
\end{figure*}

\section{Results}
We first explore the effect of varying the metallicity of the gas once it reaches  $z=22$ and a temperature of $\sim$ 1000 K, assuming the absence of UV background radiation (J$_{21}$ = 0).
Three different values of the metallicity, Z/Z$_\odot$ = -4, Z/Z$_\odot$ = -3, and Z/Z$_\odot$ = -2 are chosen. In addition, the abundances from the recently discovered CEMP star SMSS J031300.36-670839.3 \citep{Keller2014} is also investigated to probe the differences in the dynamical evolution of the halo.
In figure \ref{figure1} dynamical and chemical quantities are plotted. Increasing the metallicity, the cooling efficiency is clearly enhanced and the gas is able to reach the CMB floor temperature for Z/Z$_\odot$~=~-2.
The thermal evolution for Z/Z$_\odot$~=~-4 is mainly dominated by H$_2$ cooling and is not influenced by the small amount of metals. This result is in agreement with the critical metallicity proposed by \citet{Bromm2001} of Z/Z$_\odot$ = -3.5 and the recent results reported by \citet{Safranek2014} for atomic cooling halos. 

Once we employ the abundance pattern of SMSS J031300.36-670839.3 the evolution of the halo is very similar to the case with Z/Z$_\odot$ = -2. Indeed, the main coolant is the neutral carbon CI which regulates the whole evolution. Assuming a high metallicity is then equivalent to considering a carbon-enhanced halo. 
In figure \ref{figure1} we also report the evolution of neutral and ionized carbon (CI and CII) from where it is clear how the chemical species affect the thermal evolution. 
In our simulations we assume as initial conditions that metal species are ionized with the exception of oxygen because of its high ionization potential. The fast recombination of these species is well visible in the figure around a radius of 0.01 pc where CII declines (middle right panel) and whereas neutral carbon (CI) increases (middle left panel). 

As the initial conditions are rescaled by the metallicity the evolution of CI and CII behaves essentially in the same way but is shifted toward higher values when we increase Z/Z$_\odot$.
Due to the fact that we are not including CO, H$_2$O, and OH, the evolution of metals reaches a plateau as they are not depleted into molecules. Based on previous results, we expect that these molecules would simply provide an alternative cooling channel at high densities \citep[e.g.][]{Omukai2005}.
Oxygen and silicon species show a similar evolution. 
Changing the metallicity does not change the behavior of dynamical quantities like radial infall velocity and accretion rates, which show similar features.

In figure \ref{figure2}, the averaged projections of density, temperature, and CII fraction for different metallicities are shown. A more compact central structure and some fragmentation is already visible for metallicities higher than Z/Z$_\odot$ = -4.  The temperature is clearly lower for Z/Z$_\odot$ = -2, while Z/Z$_\odot$ = -4 presents a temperature of about 200 K in the central core comparable to the primordial case. The ionized carbon fraction decreases in the core where the temperatures are lower and recombination reactions faster.

\subsection{The effect of Far-UV radiation}
In a second series of runs here we explored the effect of varying the intensity of the radiation flux expressed in terms of J$_{21}$ (10$^{-21}$ erg s$^{-1}$ cm$^{-2}$ sr$^{-1}$ Hz$^{-1}$) that is kept below the Lyman limit ($h\nu < $ 13.6 eV). We fix the metallicity to Z/Z$_\odot$ = -3. The aim of these runs is to better understand under which conditions the halo is still able to collapse and reach the CMB temperature.
In figure \ref{figure3} the most relevant chemical species are shown as a function of radius for different values of J$_{21}$. The following important features are noted:
\begin{itemize}
	\item The H$^-$ and H$_2$ fractions are strongly suppressed by the presence of a stronger UV flux for radii grater than 0.1 pc. H$_2$ starts to form once H$^-$ starts getting abundant and formation dominates over destruction.
	\item The CII evolution is almost constant. As we start from ionized metals and considering the low ionization potential of CI, recombination is only able to form a small amount of CI, while CII is continuously pumped by the radiation. This is different with respect J$_{21}$ = 0, where recombination is the main reaction path for CII which is then quickly destroyed.
	\item SiII and SiI show a similar behavior as CII and CI. Oxygen is not reported as the flux is not able to ionize it (ionization potential of 13.61 eV) and its evolution is not affected by the changes in J$_{21}$.
\end{itemize}

Even for very different values of J$_{21}$, a very similar thermal evolution is obtained for J$_{21}$ = 1, 10, 100 at a metallicity of Z/Z$_\odot$ = -3. 
An increase in metallicity thus cancels the effect of having a very high flux (J$_{21} = 100$) allowing the halo to collapse similarly to the case without any radiation but with enhanced cooling at higher densities. This is a clear confirmation that even in the absence of H$_2$ the halo is able to collapse due to metal enrichment. 
This is even clearer from the data reported in table \ref{table1} where the virial temperature, mass, and redshift for every run has been evaluated following \citet{Barkana2001}.  Runs A and E present similar collapse parameters and a similar evolution. The presence of a higher flux at metallicity Z/Z$_\odot = -3$ allows the halo to reach a higher virial temperature before the collapse because at low density the H$_2$ cooling is suppressed. The gas should then reach a high enough density and temperature for metal cooling to be efficient. This effect causes a delay of the collapse to $z_\mathrm{vir}$~=~15 and a higher virial temperature T$_\mathrm{vir}\sim 4700$~K, allowing the metals to drop the gas temperature to even lower values due to the lower CMB temperature (40 K instead of 60 K). 
Runs B, C, and D, have similar properties.
We can therefore conclude that the main net effect of the presence of radiation is to delay the collapse and to activate a possible channel for the formation of lower mass stars. On the other hand, an increase in metallicity removes any possible effect of the radiation. This finding is in agreement both with the results reported by \citet{Jappsen2009} but also with the findings of \citet{Bromm2001} and \citet{Safranek2014}.
The critical metallicity value represents indeed a threshold for the cooling-mode only in the absence of UV flux, while it is a real threshold to induce possible fragmentation in the presence of background.
From recent studies \citep[e.g.][]{Mark2014}, it is however evident that such a radiation field will generally be present, therefore effectively requiring a critical metallicity for collapse.

\begin{figure}[!h]
\includegraphics[scale=.40]{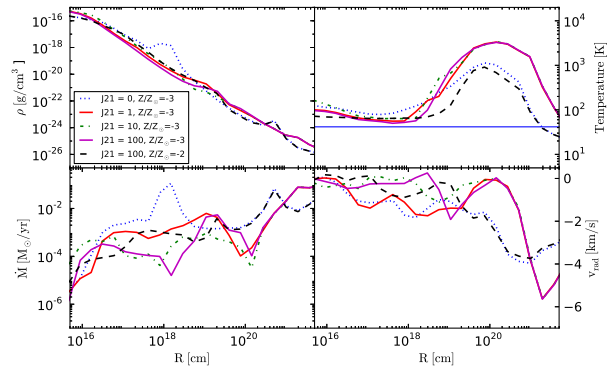}
\caption{Radial profiles of averaged density (top left), temperature (top right), accretion rate (bottom left), and infall velocity (bottom right), for same parameters as in figure \ref{figure3}. The horizontal blue line represents the CMB temperature.}\label{figure4}
\end{figure}

Finally the overall behavior of the radial infall velocities and accretion rates reported in figure \ref{figure4} is again similar for all the runs; in general stronger flux and higher metallicities produce higher infall velocities.  


%

\begin{deluxetable}{lllllll}
\tabletypesize{\scriptsize}
	\tablecaption{Simulations details: flux strength (J$_{21}$), metallicity (Z/Z$_\odot$), virial mass in M$_\odot$, redshift, and temperature of the halo. In the last column the CMB temperature T$_\mathrm{cmb}$ = 2.73(1+$z$) is also reported.}
	\tablewidth{0pt}
	\tablehead{
	\colhead{Run} & \colhead{J$_{21}$} & \colhead{Z/Z$_\odot$} & \colhead{ Mass (M$_\odot$)} & \colhead{ $z_{\mathrm{vir}}$} & \colhead{T$_{\mathrm{vir}}$ (K)} & \colhead{T$_\mathrm{cmb}$ (K)} 
	}	
	\startdata			
			A &  0 & 	-3 & 1.4$\times$10$^5$ 	& 21	& 1100  & $\sim$60\\
			B & 1     & 	 -3 & 2.0$\times$10$^6$	& 15	& 4700 &$\sim$44\\
			C & 10 & -3 & 2.0$\times$10$^6$	& 15	& 4700 &$\sim$44\\
			D & 100& -3 & 2.0$\times$10$^6$	& 15	& 4700 &$\sim$44\\			
		         E & 100& -2 & 1.4$\times$10$^5$& 21 & 1100 &$\sim$60 \\
		        \enddata \label{table1}
\end{deluxetable}
 
\section{Conclusions}
In this letter, we focused on the chemo-dynamical conditions necessary to allow a minihalo enriched by metals to undergo collapse. 
We included an accurate treatment of the microphysics employing the chemical package \textsc{krome}. A complete non-equilibrium low-metallicity network has been evolved including 16 species and 44 reactions, photochemistry, non-equilibrium individual metal line cooling, together with standard primordial thermal processes. This is the first time that such a network together with an accurate treatment of metal line cooling is employed in the 3D hydrodynamical code \textsc{enzo}. All the species have been advected to guarantee mass conservation. We consider both solar-type abundance patterns as well as pollution from type II supernovae as recently reported by \citet{Keller2014} and \citet{Cooke2014}.  Metals are injected around redshift $z = 22$ once the halo reaches a temperature of about 1000~K with a mass of 1.4$\times$10$^5$ M$_\odot$ simulating a mini-halo polluted by metals dispersed from a nearby star death. This metal-enrichment strategy is of course only an approximation as self-consistent simulations of metal-enrichment and subsequent collapse exceed the current computational capabilities \citep[see for example][]{Greif2010}. 

Different metallicities have been explored ranging from Z/Z$_\odot$ = -4 to Z/Z$_\odot$ = -2, also including the abundances pattern provided by recent observational data \citep{Keller2014}.
To better assess the physical conditions for the collapse, we included a UV radiation flux with different strengths considering a T$_*$ = 10$^4$ K soft spectrum and including H$_2$ photodissociation, H$^-$ photo-detachment, as well as self-shielding from \citet{Wolcott2011sfh}. In the presence of UV flux the halo grows until 2$\times$10$^6$ M$_\odot$ before starting to collapse at redshift 15. This allows the halo to reach lower temperatures compared to the case without radiation ($\sim$ 40 K) where collapse sets in earlier for higher temperatures of the CMB. It is then likely to produce lower mass second generation stars.
As expected, for J$_{21}$ = 0, a change in the metallicity provides a switch from H$_2$ dominated cooling to metal-line cooling for metallicities above Z/Z$_\odot$ = -3. It should be noted that the expected average intensity for far UV radiation has been recently calculated to be above J$_{21}$ = 1, even for higher redshifts \citep{Mark2014}. 

A critical metallicity according to \citet{Bromm2001} has been found between Z/Z$_\odot$ = -4 and Z/Z$_\odot$ = -3. Even in the absence of H$_2$ the halo is able to cool and collapse, while above this critical metallicity, the floor limit is given by T$_\mathrm{cmb}$. While neutral carbon is the main coolant for J$_{21}$ = 0, in the presence of radiation the gas keeps the metals ionized and CII becomes more important. The differences between different flux strengths are not very pronounced. In the case of strong flux, J$_{21}$~=~100, and higher metallicity, Z/Z$_\odot$~=~-2, the thermal evolution of the gas is very similar to the case with J$_{21}$~=~0, because the cooling from fine-structure metal transitions is very strong and the halo collapses at an earlier redshift without need of growing in mass and reach higher densities. 

We therefore conclude that with a significant amount of metals, in the presence or absence of radiation, collapse will occur. The metals will cool the gas to the temperature of the CMB, possibly induce fragmentation, and therefore
determine the mass scale of the resulting stars.
It is thus evident  that carbon-induced cooling is central during high-redshift  structure formation.

\acknowledgments
S.B. and D.R.G.S. thank for funding through the DFG priority program `The Physics of the Interstellar Medium' (project SCHL 1964/1-1). D.R.G.S. and M.L. thank for funding via the SFB 963/1 on "Astrophysical Flow Instabilities and Turbulence" (project A12). The plot of this paper have been obtained by using the $\mathrm{YT}$ tool \citep{Turk2011a}. The simulations have been performed on the HLRN cluster under the project nip00035.


\end{document}